\documentclass[rnote]{aa}
\usepackage{natbib,amssymb,graphicx,graphics}
\usepackage{subfigure}
\usepackage{txfonts}
\newcommand{\excs}{\extracolsep{\fill}}

\begin{document}
\title{Coronagraphic near-IR photometry of AB\,Dor C}
\titlerunning{Photometry of AB\,Dor C}

\author{A. Boccaletti\inst{1}  \and G. Chauvin\inst{2}  \and P. Baudoz\inst{1}  \and J.-L. Beuzit\inst{2}}
\institute{LESIA, Observatoire de Paris-Meudon F-92195, Meudon, France \\
\email{anthony.boccaletti@obspm.fr, pierre.baudoz@obspm.fr} 
\and Laboratoire d'Astrophysique de l'Observatoire de Grenoble, F- Grenoble, France \\
\email{gael.chauvin@obs.ujf-grenoble.fr, jean-luc.beuzit@obs.ujf-grenoble.fr} }

\offprints{A. Boccaletti}
\date{Received ; accepted }

\abstract{
Observations of low-mass companions for which the dynamical masses are well constrained help to improve the calibration of evolutionary models. Such observations thereby provide more confidence in the estimation of the mass of a companion using the photometric methods expected for the next generation of planet finder instruments. 
  }  
  {The commissioning of  a new coronagraph at  the Very Large Telescope (VLT) was  the occasion to
  test  the performance  of  this  technique on  the  well-known  object
  AB\,Dor A and its 0.09$M_\odot$  companion AB\,Dor C. The purpose of
  this paper is to refine  the photometric analysis on this object and
  to  provide an  accurate photometric  error budget.}{In  addition to
  coronagraphy, we calibrated the residual  stellar  halo with  a
  reference star. We used standard  techniques for photometric extraction. }
  {The  companion AB\,Dor C  is easily detected at  $0.185~\!''$ from
  the primary  star, and its magnitudes in  $H$ and $Ks$ are in agreement with an M$5.5$ object, 
  as already known from spectroscopic observations. However, these new measurements 
  make the earlier $J$-band photometry less reliable. Finally, the comparison with evolutionary models supports an age of $(75\pm25)$~Myr, contrary to previous analyses. 
  These   observations 
  demonstrate  that coronagraphic observations  can be  more efficient
  than direct imaging, not only to improve contrast, but also to provide
  a better photometric estimation as long as a good calibration of the
  stellar halo is achieved.}{}


\maketitle

\section{Introduction}

Since the last  decade, the improvement of high  angular resolution on
large telescopes  has made possible the discovery  of faint companions
with  masses   close  to  the   planetary  mass  regime,   2M\,1207  B
\citep{Chauvin05};    DH\,Tau B \citep{Itoh05};    GQ\,Lup    B
\citep{Neuhauser05};  AB\,Pic  B  \citep{Chauvin05a}; and  CHXR\,73  B
\citep{Luhman06b}. The  precise determination of  the mass of  most of
these  companions  is presently  not  possible  as  long as  dynamical
measurements are  missing. However, an  estimation of the mass  can be
obtained  from   photometric  measurements  via   evolutionary  models
\citep{Burrows97,  Chabrier00}. Calibration  of these  models  on very
low-mass objects,  for which  the mass is  known from  other techniques
(radial velocity,  astrometry), is  highly desirable to  prepare future
instruments.   The instruments SPHERE \citep{Beuzit06}  and   GPI  \citep{Macintosh06}   will
precisely  use  broadband  differential imaging  or  low-resolution
spectroscopy   to  carry  out   statistical  analysis   on  extrasolar
planets. Hence,  accurate mass estimation is critical  in this context
and a strong effort has been made to refine atmospheric models of giant
planets and brown dwarfs. 

Young, nearby associations  are well suited to the identification
and the  follow-up observations  of young dynamical  mass calibrators,
such  as  the  tight  binaries  HD\,98800 \citep{Boden05}  and
TWA5\,Aab \citep{Konopacky07} of the TW Hydrae association.  Very
recently, much attention has been paid to the hierachical quadruple
system AB  Dor, a member of  the eponymous comoving group  identified by
\citet{Zuckerman04}.  The brightest component AB\,Dor A was first
known as a variable star featuring variation of 0.09 mag in the V band
and flares of 0.05 mag  near its maximum \citep{Innis85}.  It was then
recognized  as a  rapidly-rotating spotted  star  and was  intensively
studied  as  such  in  the  1980s.   Accurate  parallax  obtained  with
Hipparcos ($\pi=(6.92\pm0.54)$~mas, $d=(14.9\pm 0.12)$~pc; \citet{Perryman97}) 
allowed \citet{Wichmann98} to  derive a spectral type of K1,
while it was previously thought to be a post-T Tauri star.  

At $9.0~\!''$ North, the physical companion AB\,Dor B
\citep{Lim93, Guirado06}  is resolved   as  a   tight
($\Delta=0.070~\!''$) binary by \citet{Close05}.
The object AB\,Dor  C is  the fourth  component of  this young  quadruple system,
discovered thanks to the reflex motion induced on AB\,Dor A detected with Very Long Baseline Interferometry
and  Hipparcos observations \citep{Guirado97}.  
\citet{Close05} have refined  the   mass  estimation  of   AB\,Dor C  to $(0.090\pm0.005)~M_\odot$ (confirmed later by \citealt{Guirado06}).   
A first  attempt  to image  this close  and
low-mass companion with ADONIS at the 3.6m telescope of La Silla (ESO)
was   unsuccessful   due   to   the   lack   of   angular  resolution
\citep{Boccaletti01}.  \citet{Close05}  finally resolved AB\,Dor  C at $0.156~\!''$
from A using VLT/NACO.  They  measured the near IR absolute magnitudes
(see Tab. \ref{tab:photo}) and  a spectral type M$8\pm1$ for  this faint companion.
The $J$-  and $H$- brightnesses and the  deduced effective temperature
were found to be inconsistent with evolutionary models, considering an
age  estimate   of  $50_{-20}^{+30}$~Myr  based   on  different  youth
indicators.  In conclusion, \citet{Close05} suggested that  theoretical
models are  actually underestimating the  mass in the young age and low-mass  regime.
\begin{figure*}[t]
\centerline{\includegraphics[width=6cm]{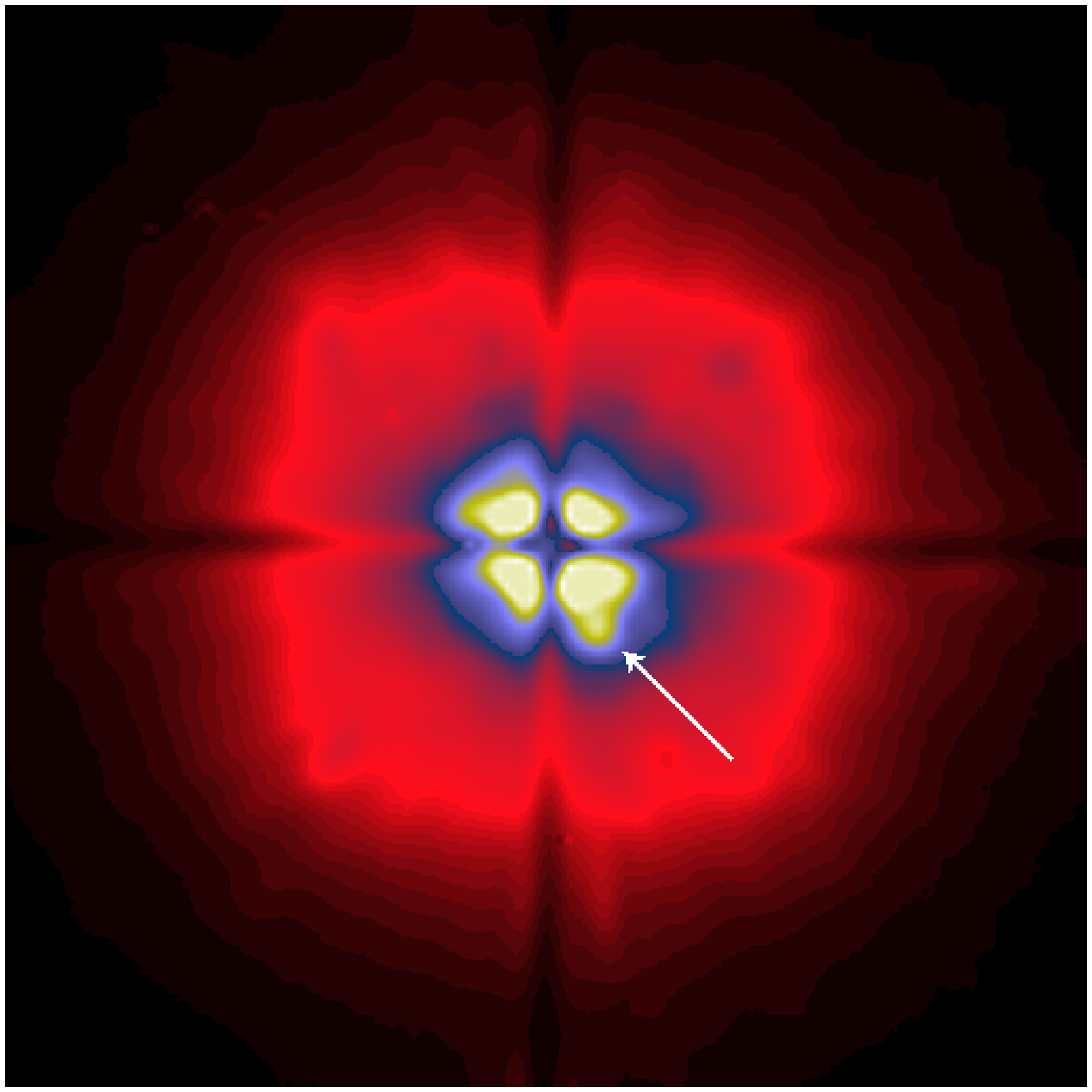}
\includegraphics[width=6cm]{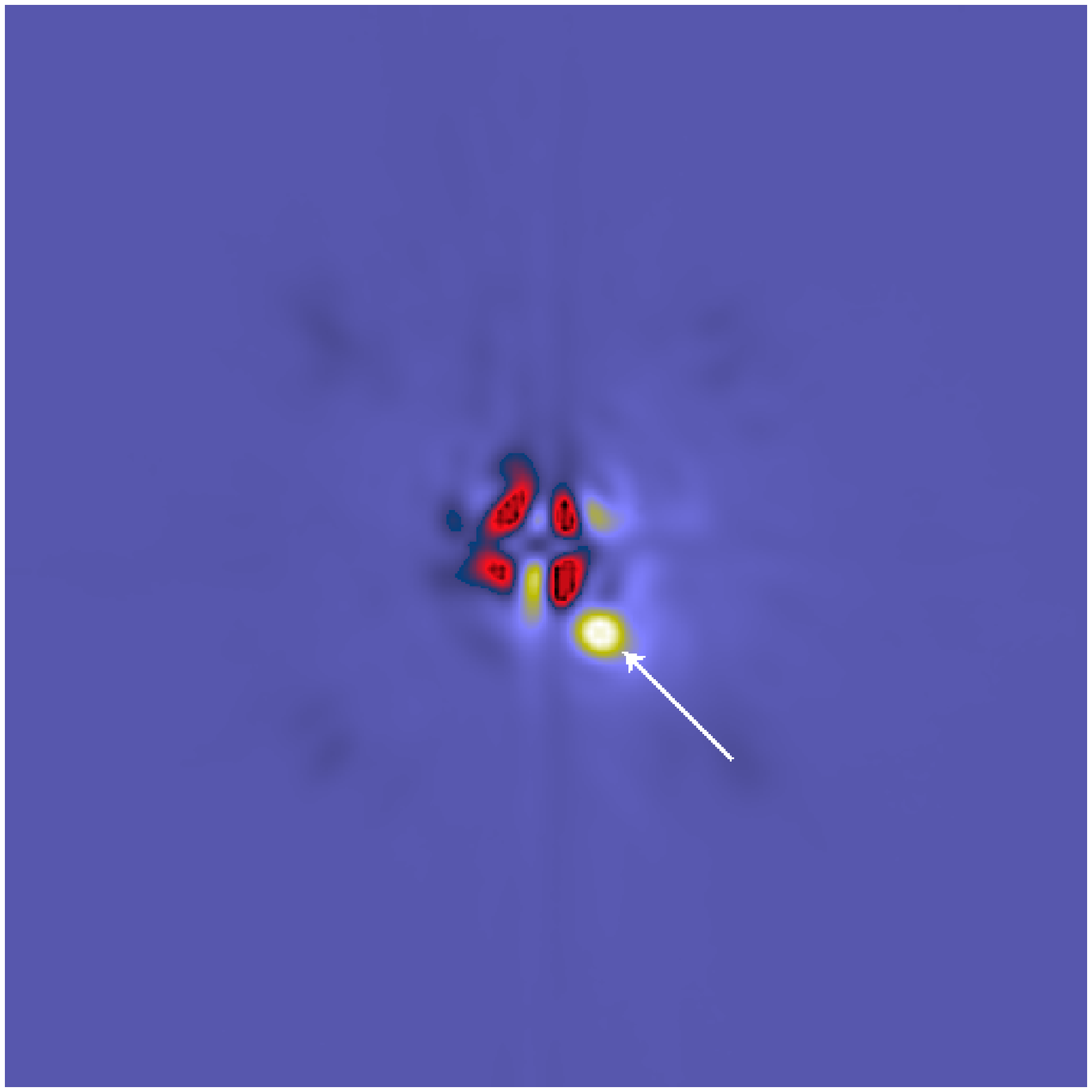}}
\caption[]{Coronagraphic image of AB\,Dor obtained in the $Ks$ filter
  (\textit{Left}) and the same image subtracted with  a reference star (\textit{Right}). The field
  of view is $2~\!''$. North is up, East is left. Arbitrary false colors are
  intended to enhance the companion visibility.} 
\label{imaKs}
\end{figure*}

\citet{Luhman06} revised the JHK photometry uncertainties and
the spectral type estimation of \citet{Close05} based on the same data set.  Using,
in addition,  a different age estimate  of $75-150$~Myr for  the AB\,Dor
association  \citep{Luhman05}, they  concluded  that there  was
currently no disagreement between models and data.  Recent VLT/SINFONI
spectra in  HK-bands enable \citet{Close07} to derive  a spectral type
M5.5$\pm$1,  which  confirms the  conclusions  of  \citet{Luhman06}.  This  prolific
system  illustrates   the  difficulty  of   testing  evolutionary  model
predictions  without accurate  observables (effective  temperature and
luminosity) and robust age estimate.

In 2006,  a proposal to combine the  Simultaneous Differential Imaging
(SDI)  mode  of  NACO (the Nasmyth Adaptive Optics System and Near-Infrared Imager and Spectrograph) 
with   a  4  Quadrant  Phase  Mask  coronagraph
\citep{Rouan00} was  approved by European Southern Observatory (ESO). During the  commissioning run we
collected data on AB\,Dor A and C (Sect. \ref{sec:obs}) and here, we
present the results of our $H$ and $Ks$ photometric analysis (Sect.
\ref{sec:photo})  together with  a detailed  estimation of  error bars
(Sect. \ref{sec:error}). Results are discussed in Sect. \ref{sec:discu}.

\section{Observations}
\label{sec:obs}

We carried  out observations  as  part of  a  commissioning run  on
February 16th,  2007 at  ESO/Paranal.  The AO-assisted  near-IR camera
NAOS-CONICA named NACO \citep{Rousset03, Lenzen03} was equipped with a
new set of  two 4 Quadrant Phase Masks  \citep{Rouan00} to replace the
old   one  \citep{Boccaletti04}.   These  two   masks   are  operating
respectively in  the $Ks$ and  $H$ bands, the latter  being compatible
with  the  Simultaneous  Differential  Imager (SDI)  provided  by  the
University  of Arizona  and  the Max  Planck  Institute of  Heidelberg
\citep{Lenzen04}.

We observed AB\,Dor (V=6.93,  H=4.845, K=4.686, Sp=K1III)  with the 4
Quadrant Phase Mask  (4QPM) in two filters. We  obtained 600s (DIT=1s,
NDIT=100, Ncycle=6)  of data in  the $Ks$ band ($\lambda=2.18  \mu m$,
$\Delta\lambda= 0.35\mu  m$) on the  target and a  similar integration
time on a reference star  (HD\,41371) chosen with the same visible and
IR  fluxes (V=7.10,  K=4.724, Sp=K0III)  and observed at  the  same parallactic
angle. This  optimal observing  strategy preserves the  orientation of
the telescope  pupil with  respect to NACO,  and therefore,  reduces the
differential  aberrations between the  star and  its reference  as the
matching of spider spikes in  the two images. In addition, we observed AB\,Dor 
with SDI (the  4QPM being  installed in  the beam)  for 936s
(DIT=8s,  NDIT=13,   Ncycle=9).  Instead  of  a   reference  star,  the
calibration  of  the  speckled  halo  is  obtained  simultaneously  in
different    filters   ($\lambda=1.575,    1.600,   1.625    \mu   m$,
$\Delta\lambda=0.025 \mu  m$). However, a second  level of calibration
is required to  reduce the impact of differential  aberrations and the
speckle   chromaticity   (as  the   phase   varies  with   wavelength,
\citealt{Marois00}). For  this purpose, we obtained two observations  with the field
of view rotated by  $60^\circ$. Differential aberrations
are assumed static in this case.

Coronagraphic observations with NACO  are preceded with an acquisition
template that provides an out-of-mask PSF to be used as a photometric
reference. However, this reference  is obtained with a different setup
than  coronagraphic frames. Because AB\,Dor  is a  bright object,  a Neutral
Density (ND)  is needed  to avoid detector  saturation (for  $Ks$ band
data only), and as default a full aperture stop (Full) is used instead
of   a   stopped    aperture   (Full\_uszd)   as   for   coronagraphic
templates. Photometric  measurements have  to be corrected  from these
values.

Seeing conditions and AO correction were good. The average seeing was
$(0.91\pm0.13)"$ for the $Ks$ observing block and $(0.78\pm0.12)"$ for the
SDI data  while the coherent  energy measured from residual  slopes of
the AO system was, respectively, $(46\pm5)\%$ and $(52\pm4)\%$.

\section{Data reduction and photometric measurements}
\label{sec:photo}

\begin{figure*}[t]
\centerline{\includegraphics[width=6cm]{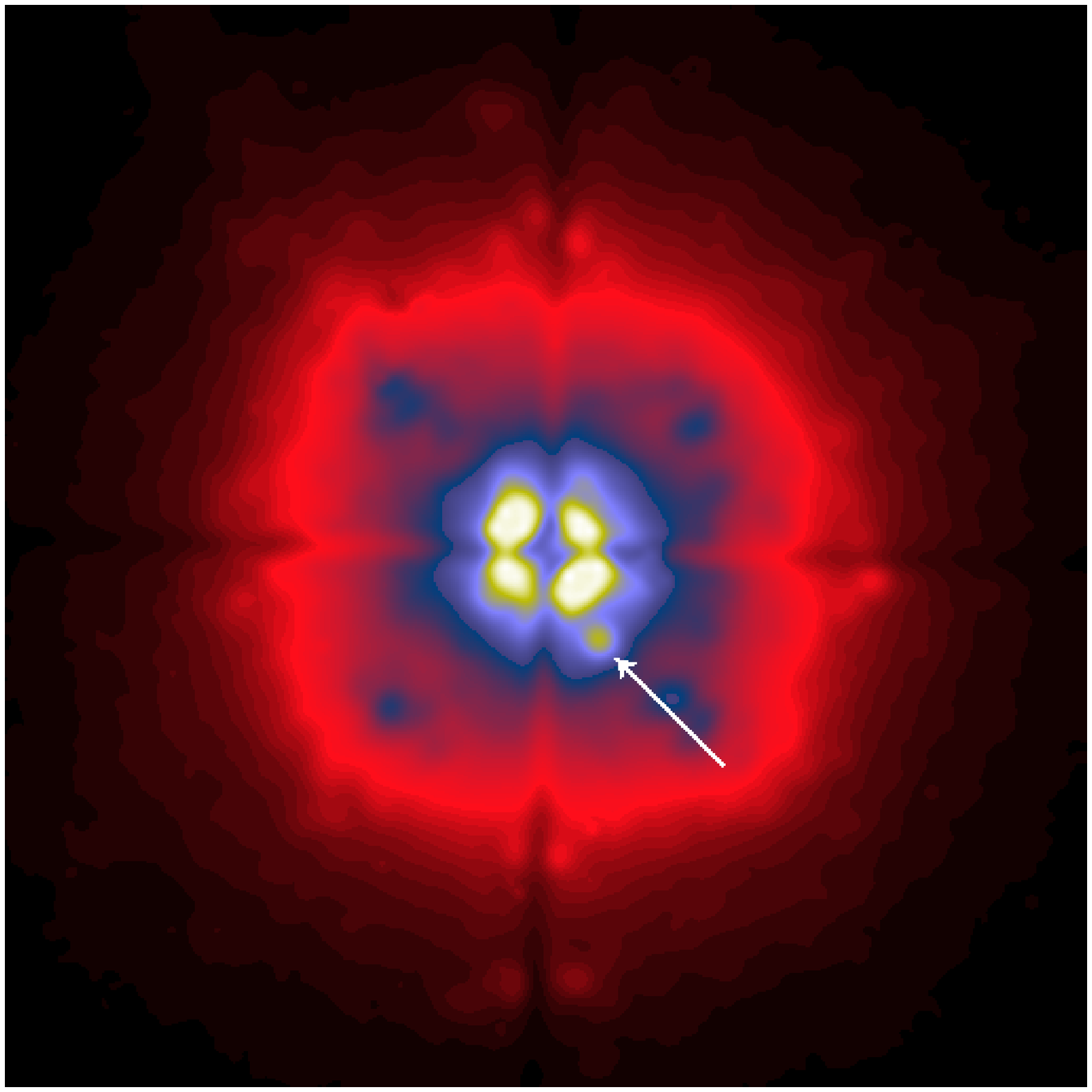}
\includegraphics[width=6cm]{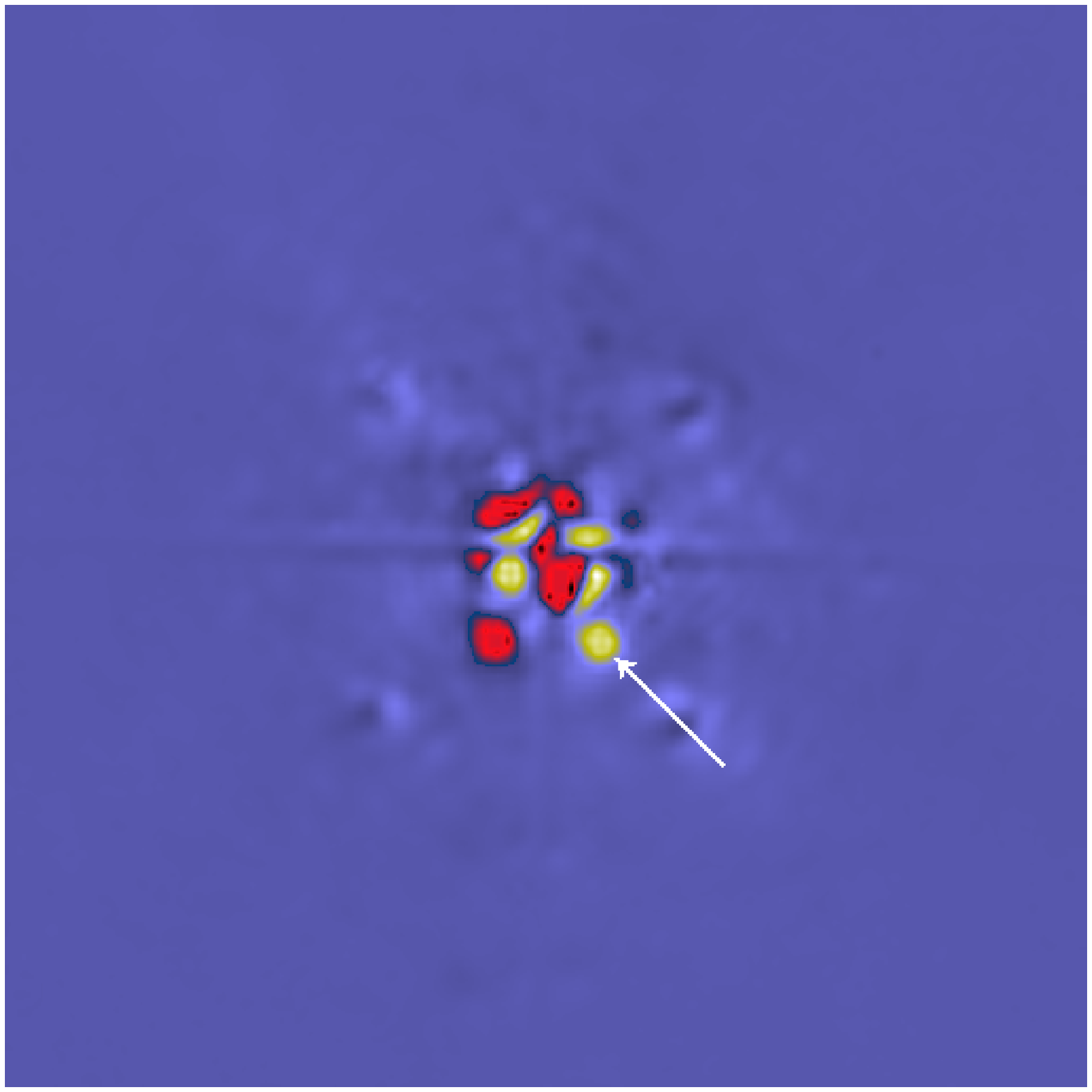}
\includegraphics[width=6cm]{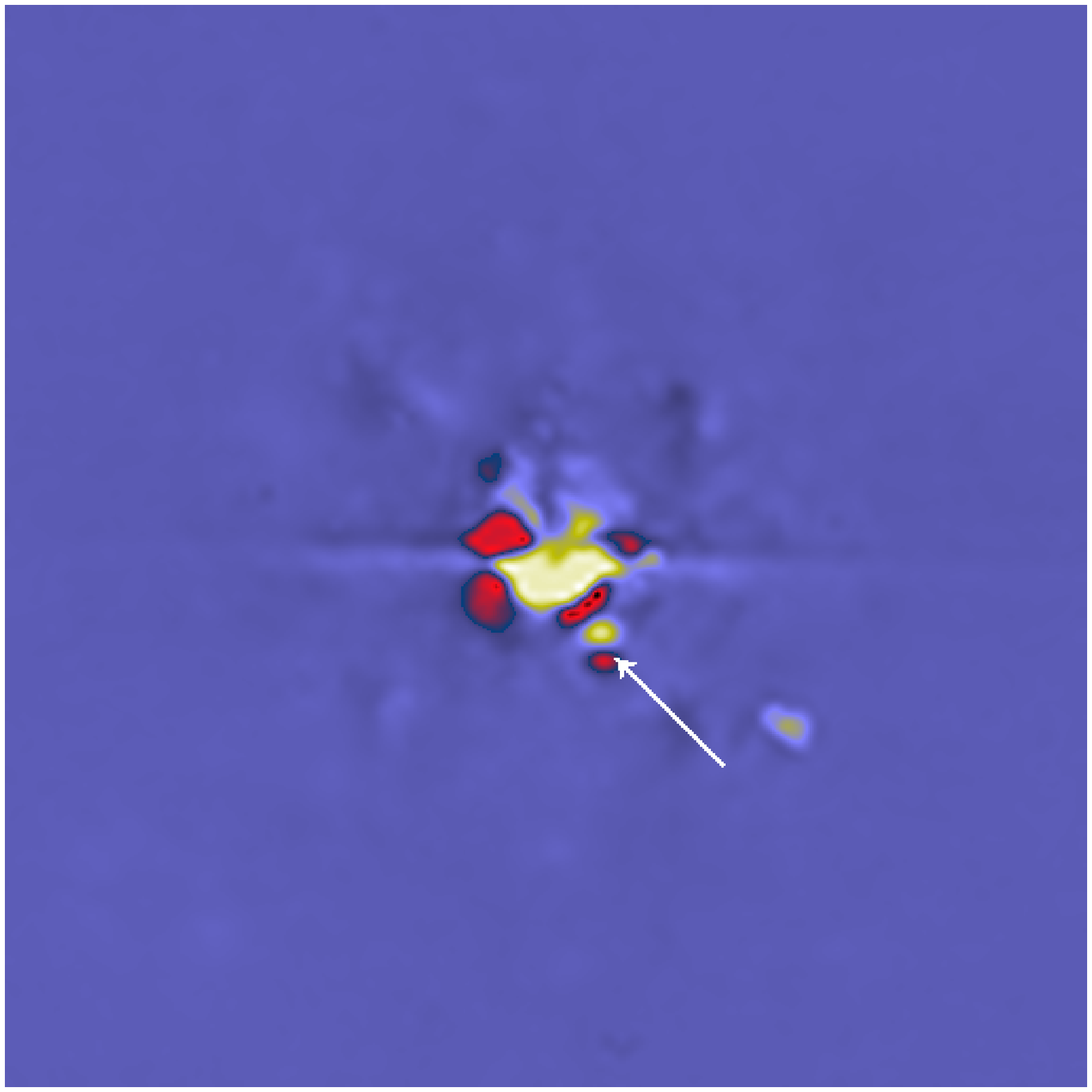}}
\caption[]{Coronagraphic image of AB\,Dor obtained with SDI (\textit{Left}) and
  the same image subtracted with a field-rotated AB\,Dor image (\textit{Middle}) compared
  to  the 2  wavelengths subtraction  (\textit{Right}). Subtracted frames show a positive image of the companion (yellow) and a negative one (red). The SDI frame is not exploitable because of the small angular separation making the positive and negative component self-subtracting. The field  of view  is
  $2~\!''$. North is  up, East is left. Arbitrary  false colors are intended
  to enhance the companion visibility.} 
\label{imaH}
\end{figure*}

We processed the data with standard reduction routines to
correct  for bad  pixels, flat  field uniformity  and to  subtract an
average  sky  background.  We obtained flat fields without  the
coronagraph in the  beam, although it would have  been better to reduce
the presence of dust particles on the substrate. However, the field of
view slightly  drifted in  front of  the detector  plane  (as the
instrument  rotates   at  the  Nasmyth  focus),   making  the  precise
registration of dust features on the coronagraph substrate impossible.

We corrected  for the ND  attenuation on the  $Ks$ data (a  factor of 89) and 
for the difference of pupil stop (a factor of 0.808 estimated from geometrical assumptions) 
to build a  master normalized PSF.  The individual coronagraphic
cycles were then co-added for  the star and the reference separately so
as to provide two images. Recentering was not required. At this stage,
the   companion   is  yet visible  (Fig.   \ref{imaKs}   and
Fig. \ref{imaH},  \textit{Left}), but we needed the  subtraction of the reference  star
to  remove the PSF halo  and to accurately measure the
photometry. To estimate the intensity factor between the coronagraphic
image  of the  star and  that of  the reference, we considered  several  methods: 
total  intensity ratio,  balance of positive  and negative fluxes, balance 
of positive and  negative pixels, and  minimization of
the total residual intensity. We adopt the last method as a baseline paying
attention to the presence of  the companion to  avoid a bias. We  found an
intensity  factor of  $1.04 \pm  0.01$ with  small  dispersion between
methods    and    produced    a   subtracted    coronagraphic    image
(Fig. \ref{imaKs}, \textit{Right}).

Once the stellar  contribution is removed at the  companion location a
thorough estimation of the  companion intensity becomes possible. Here
also,  we  compared  several  methods like  aperture  photometry,  PSF
fitting with a 2D gaussian,  PSF to companion maximum intensity ratio,
and minimization of the residual after PSF subtraction (on the companion). In the two first
cases, the companion intensity is  integrated in a limited aperture (a
few pixels in radius) and a correction of -0.1mag is needed to account
for  the   intensity  in  the  PSF  wings   (comparison  of  encircled
energy).  As  a baseline,  we  used  the  minimization to  derive  the
average photometry.

In the particular case of SDI data, two wavelengths subtraction is
inappropriate  since AB\,Dor C  does not  contain methane  and the
bifurcation point  (the distance at  which the companion image  in the
rescaled  frame falls  at  different pixels  than  in the  un-rescaled
frame,  \citealt{Thatte07}) is located  at 3.2$~\!''$.   However, observational
procedure with SDI  requires the acquisition of a  field-rotated image
of  the same star to correct  for the  differential static
aberrations  and  the  chromatic   aberrations  inherent  to  the  SDI
technique. Therefore, we measured $H$ band flux of AB\,Dor C with the
field-rotated  image as a reference  star both in a  single SDI filter
($\Delta\lambda=0.025 \mu m$) and  with the combination of all filters
($\Delta\lambda=0.075  \mu m$) to approach  the  broadband
magnitude. Hence, two values are  provided for the $H$ band magnitudes
that we note: $m_{H_{1\lambda}}$ and $m_{H_{3\lambda}}$.

As a  result, we found  the following magnitude differences  : $\Delta
m_{H_{1\lambda}}=4.71$,  $\Delta  m_{H_{3\lambda}}=4.62$  and  $\Delta
m_{Ks}=4.56$. The angular separation is $0.185~\!''$ consistent with the
orbital solution presented in \citet{Nielsen05}. However, it was not the goal of this paper to discuss the astrometry of the companion since it was already characterized from Hipparcos data and confirmed by the aforesaid papers. In addition, the presence of the coronagraph makes the estimation of the astrometry less accurate unless appropriate techniques are considered \citep{Marois06}.

\section{Photometric errors}
\label{sec:error}

\begin{table}[t]
\caption{Photometric uncertainties in magnitude.} 
\label{tab:errors}
\centering  
\begin{tabular}{l c c }  \hline\hline 
sources of error		&	$H$		&	$Ks$ 	\\ 	\hline
pupil stop				& 0.05		& 0.05		\\
ND					&  -			& 0.06		\\
intensity factor 		 	& 0.05		& 0.04		\\
method for extraction	& 0.13		& 0.13		\\
aperture size			& 0.04		& 0.01		\\
distance				& 0.03		& 0.03		\\
2MASS				& 0.03		& 0.02		\\
filters conversion               & 0.03 		&0.01 		\\ \hline 
total quadratic error		& 0.16 		& 0.16		\\	\hline
\end{tabular} 
\end{table} 
\begin{table*}[t]
\caption{Photometry  of  AB\,Dor,  A  being  the  primary  and  C  the
  companion converted into the 2MASS system.} 
\label{tab:photo}
\centering  
\begin{tabular*}{\textwidth}{@{\excs}llllllll}
\hline\hline\noalign{\smallskip}
UT Date & M$_{J}$ 	& M$_{H}$ 	& M$_{Ks}$ 	& $\Delta$ 	& SpT 	& $T_{\rm{eff}}$ 	& References \\
   	       & (mag) 	& (mag) 		& (mag) 		& (mas) 		& 		& (K)			 	& \\
\noalign{\smallskip}\hline\noalign{\smallskip}
04.02.2004 & $9.89^{+0.19}_{-0.24}$ & $9.17^{+0.13}_{-0.15}$& $8.58^{+0.12}_{-0.15}$ & 156 & M$8\pm1$ & $2600_{-150}^{+150}$ & \citet{Close05}\\
           &   $9.85^{+0.32}_{-0.46}$   &   $9.31^{+0.24}_{-0.31}$&    $8.92^{+0.28}_{-0.37}$   &   156   &   M$6\pm1$   &$2840_{-120}^{+170}$ & \citet{Luhman06}\\
07.01.2005 & & & $8.63\pm0.17$& 220 & & &  \citet{Close07} \\
24.01.2006 & & & & 200 & M$5.5\pm1$ & $2925_{-140}^{+170}$&  \citet{Thatte07}, \citet{Close07}\\
16.02.2007 & & $8.71\pm0.16$ & $8.38\pm0.16$ & 185& & & this work \\
16.02.2007 & & $8.64\pm0.16$ &  & & & & this work \\
\noalign{\smallskip}\hline
\end{tabular*} 
\end{table*} 

We identified several  sources   of  errors  in  the  photometric
extraction that we analyze  in this section to derive error bars:

\begin{itemize}
\item[-] The  pupil Lyot stop correction made  on the PSF  flux. As mentioned,  
a  geometrical  comparison  of  the  Full  aperture  and  the
  Full\_uszd aperture  (undersized by 10\%)  leads to a  correction of
  0.808. However,  a photometric measurement obtained in  a previous  observing run
  suggests  a  value  of  0.775.   Then,  an  uncertainty  of  4\%  was
  considered. 
\smallskip
\item[-] The  ND correction  made on  the PSF flux  (only for  data in
  $Ks$). Our measurement in the Ks band yields an attenuation factor of
  $89\pm3.6$. 
\smallskip
\item[-] Intensity factor between  the coronagraphic image of the star
  and that of the reference. The  variety of method we used to measure
  this parameter is providing a  good estimate of the uncertainty. The
  precision achieved is 1\%. 
\smallskip
\item[-] Photometric extraction. Here  again, we used a variety of methods
  to  consolidate the  result. However, we  identified it  as the
  major  source  of uncertainty  in  our  measurement. The  dispersion
  between methods is of about 0.13mag. 
\smallskip
\item[-]  Aperture   size,  when   aperture  photometry  is   used  for
  photometric extraction. The radius  of the aperture photometric mask
  is set  to $1.22 \lambda/D_{lyot}$,  $D_{lyot}$ being the diameter  of the
Lyot  stop  instead of  that  of the  telescope  (7.2m instead  of 8m).  A
  variation of 1 pixel on this radius provides the error bar. Although aperture photometry is not used to extract the photometry (see Sect. \ref{sec:photo}) it actually enters in the error term "method of extraction", but has a wavelength dependence according to the PSF sampling.
\smallskip
\item[-] The star distance which is known to an accuracy of 0.12 pc.
\smallskip
\item[-]  The  2MASS  photometric  uncertainty for  the  primary  star
  magnitude. 
\item[-]  The  conversion between systems of magnitude (NACO to 2MASS to CIT). 
\end{itemize}

Table \ref{tab:errors} gives the correspondence of these uncertainties
in magnitude.  To  derive the AB\,Dor C  absolute magnitudes in
the $H$ and $Ks$ 2MASS system, we have estimated the NACO to 2MASS filters
transformations based on the  primary and secondary spectral types and
the  filter  transmission  curves.   The  transformations  found  are
$(-0.02\pm0.03)$mag,   $(-0.04\pm0.03)$mag,   and   $(0.00\pm  0.01)$mag   in
$H_{1\lambda}$, $H_{3\lambda}$, and $Ks$ respectively.  Assuming errors
are added  quadratically and a  distance of $d=(14.9\pm 0.12)$~pc,
the new $H-$ and $Ks-$ absolute  magnitudes of AB Dor C were obtained with
uncertainties  and are  given with  the results  of previous  works in
Tab. \ref{tab:photo}. 

\section{Discussion}
\label{sec:discu}
\begin{figure}[t]
\includegraphics[width=8.5cm]{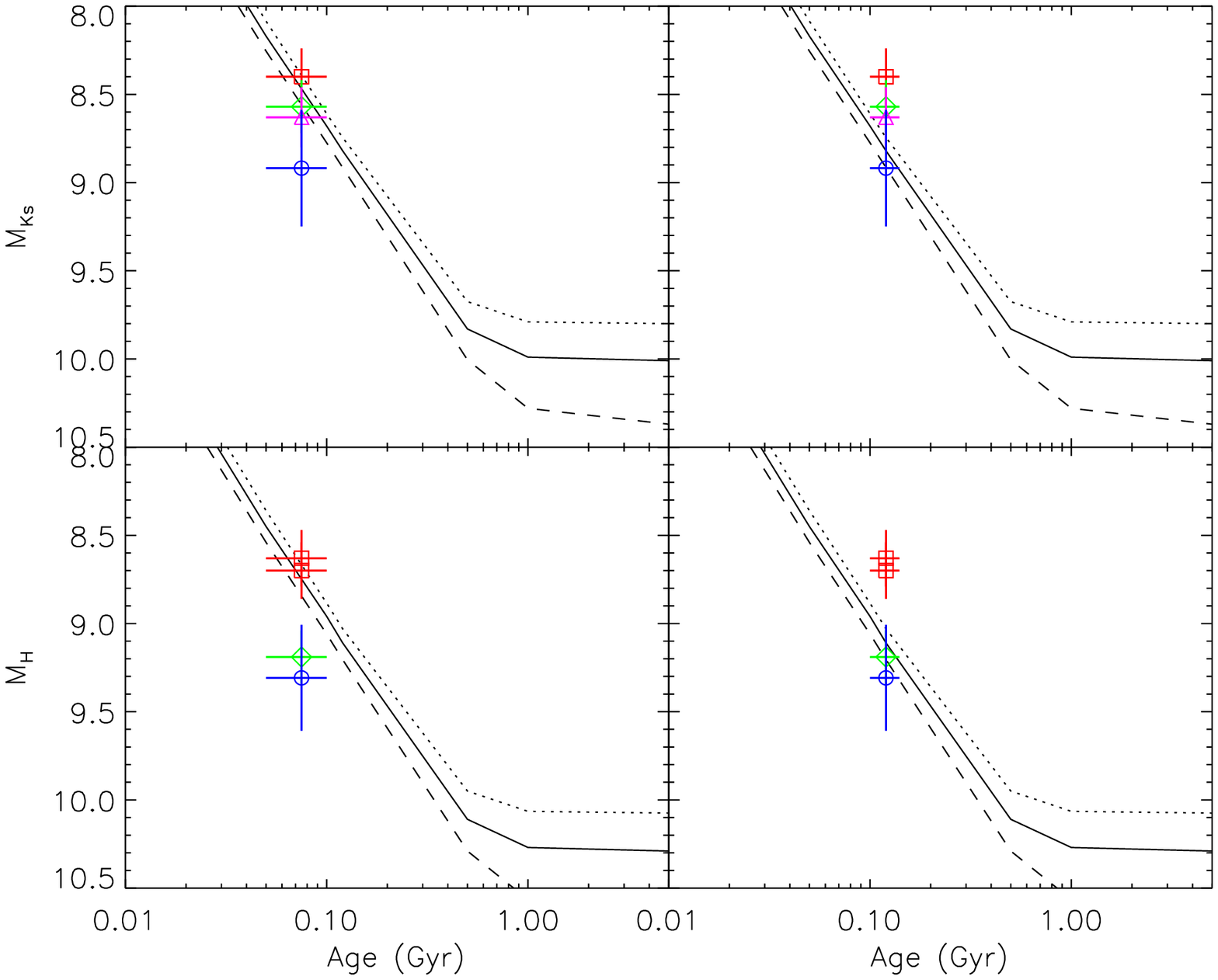}
\caption[]{Interpolated evolutionary  tracks of \citet{Chabrier00} for
  0.085 (dashed), 0.090 (solid) and 0.095 (dotted) solar masses in the
  range 0.01-10  Gyr compared  with our photometric  measurements (red
  box)  those   of  \citet{Close05}  (green  diamond) and \citet{Close07}  (purple triangle)  and  those  of
  \citet{Luhman06} (blue circle). Upper and lower plots are respectively for $Ks$  and $H$ photometry (in the CIT system of magnitudes).
  The left plots are for the first age scenario ($75\pm25$~Myr) and the right plots for the second ($120\pm20$~Myr).  } 
\label{fig:tracks}
\end{figure}

Our  new $H$  and  $Ks$ measurements  of  AB\,Dor  C and  the
derived  $H-Ks = 0.29\pm0.23$  color are  consistent with  the recent
M$5.5\pm1$ spectral  type estimation of  \citet{Close07} (using spectroscopic analysis rather than photometry).  These
new measurements also weaken  the reliability of the $J$-band photometry
of \citet{Close05} and  \citet{Luhman06} obtained  in poor
atmospheric conditions  and that  "may be systematically  too faint''
\citep{Close07}. The $J-Ks$  would be indeed even redder than before
and  surprisingly  red  compared  to  M5.5 objects  of  the  Pleiades.
Although not  impossible, the existence of  an IR excess for  AB\,Dor C
seems  to be  unlikely as  AB\,Dor A  (K1), Ba  (M$3.5\pm1.5$), and  Bb
(M$4.5\pm1.5$) do not present abnormal ($J-Ks$) colors.

Our new $H$ and $Ks$  absolute magnitudes of AB\,Dor C (color
corrected again between  the 2MASS and CIT systems)  are compared with
previous   measurements  and   over-plotted   on  evolutionary   model
predictions  of \citet{Chabrier00}  for a  dynamical mass  of $(0.09\pm
0.005)M_\odot$ (Guirado  et al. 2006).  For the  following analysis, we
have  considered  two  scenarii  of   ages  for  the  AB\,Dor  system,
$(75\pm25)$~Myr  and  $(120\pm20)$~Myr, as  the  determination  of  this
fundamental parameter is still debated.\\

The first age estimate  of $\sim50$~Myr was given by \citet{Zuckerman04}
  for the  whole  AB\,Dor  association.  We used multiple  youth
indicators, such  as  H$_\alpha$  emission, strong  lithium
$6708~\AA$  absorption,  large $v$\,sin$i$,  large  X-ray  flux, and  a
location above the main sequence  of a color ($V-K$) magnitude (M$_K$)
diagram  for three  mid-M  type  members.  \citet{Close05}  and
\citet{Nielsen05}  estimated a more precise age range to $50_{-20}^{+50}$~Myr
based on  comparisons with  other young clusters  properties.  Similar
analysis led \citet{Lopez06} to confirm this age estimate.
More recently, \citet{Janson07} focused their age-dating criteria
on the  the two M dwarfs AB\,Dor Ba and Bb,  comparing their effective
temperatures  to evolutionary track  predictions, which  are relatively
robust for low-mass stars.  However, uncorrect spectral types for both
components led  them to  over constrain the  upper age limit  for this
system.   We show model predictions on Fig. \ref{fig:tracks}  (\textit{Left}), and
compared to our measurements for an age of $(75\pm25)$~Myr. Contrary to
the  measurements of \citet{Close05} and  \citet{Luhman06}
obtained in less favorable conditions (smaller angular separation and lower AO
correction),  our   results  are   clearly  in  good   agreement  with
evolutionary  tracks in  both  $H$  and $Ks$  and  would confirm  the
validity of evolutionary models in this range of masses and ages.

Using comparable age-dating  indicators, \citet{Luhman05}
derived a  second estimate of  75-150~Myr for the AB\,Dor association.
Moreover, like \citet{Innis86}, they have  suggested that the AB\,Dor  
association might  belong  to the  Pleiades supercluster  sharing
common kinematic origin. Based  on the 3D dynamical evolution analysis
of  the AB\,Dor association  and the  Pleiades, \citet{Ortega07}
obtained a  similar age of $(120\pm20)$~Myr consistent  with both groups
being coeval.   If we now consider this age-scenario shown  on Fig. \ref{fig:tracks}
(\textit{Right}), our  measurements are not well fitted  in both bands
by the model predictions. The model predictions actually underestimate
the observed luminosity of AB\,Dor C by a factor of $\sim1.6$ ($0.5$~mag),
and thus overestimate its mass (opposite of what was found by \citealt{Close05}).

Future   age   characterization   and  confirmation of our
photometric results should help  to draw a robust conclusion to estimate
the  accuracy of the model predictions  in this  range of  ages and
masses. More  interestingly, the  results presented here  evidence the
ability  of  broadband  photometry  combined  with  coronagraphy  to
retrieve the actual companion mass, providing optical quality is met to
allow the use  of a coronagraph.  This conclusion  is important in the
context of  planet finder  instruments like SPHERE and GPI,  using differential
imaging observations of some planetary spectral features and for which
planetary  masses will  be  assessed via  the  comparison of  spectral
contrasts and evolutionary models.

\begin{acknowledgements} 
We would like to thanks the ESO team at the telescope
who support the commissioning of the coronagraph and also R. Lenzen 
for the installation of the masks inside CONICA and for his suggestions as a referee.
\end{acknowledgements} 

\bibliographystyle{aa}
\bibliography{9294ms}
\end{document}